\documentclass[12pt]{iopart}

\usepackage{multirow}
\usepackage{array}
\usepackage{hhline}
\usepackage{makecell}
\expandafter\let\csname equation*\endcsname\relax

\expandafter\let\csname endequation*\endcsname\relax
\usepackage{amsmath}
\usepackage{graphicx}
\usepackage{cite}
\usepackage{xcolor}
\usepackage{color,soul}

\DeclareUnicodeCharacter{2212}{-}

\begin{document}

\title[Deep Learning Matches Humans in Fetal US Video Measurements]{Deep Learning Fetal Ultrasound Video Model Match Human Observers in Biometric Measurements}

\author{Szymon Płotka$^{1, 2, 3}$, Adam Klasa$^3$, Aneta Lisowska$^{1, 4}$, Joanna Seliga-Siwecka$^5$, Michał Lipa$^6$, Tomasz Trzciński$^{2, 7}$ \& Arkadiusz Sitek$^1$}
\address{$^1$ Sano Centre for Computational Medicine, Czarnowiejska 36, 30-054 Cracow, Poland}
\address{$^2$ Faculty of Electronics and Information Technology, Warsaw University of Technology, Nowowiejska 15/19, 00-665 Warsaw, Poland}
\address{$^3$ Fetai Health Ltd., Warsaw, Poland}
\address{$^4$ Poznan University of Technology, Piotrowo 3, 60-965 Poznan, Poland}
\address{$^5$ Medical University of Warsaw, Karowa 2, 00-312 Warsaw, Poland}
\address{$^6$ 1st Department of Obstetrics and Gynecology, Medical University of Warsaw, Plac Starynkiewicza 1/3, 02-015 Warsaw, Poland}
\address{$^7$ Jagiellonian University, Prof. Stanisława Łojosiewicza 6, 30-348 Cracow, Poland}
\ead{s.plotka@sanoscience.org}
\vspace{10pt}

\begin{abstract}
\textbf{Objective:} This work investigates the use of deep convolutional neural networks (CNN) to automatically perform measurements of fetal body parts, including head circumference, biparietal diameter, abdominal circumference and femur length, and to estimate gestational age and fetal weight using fetal ultrasound videos. 
\newline
\textbf{Approach:}
We developed a novel multi-task CNN-based spatio-temporal fetal US feature extraction and standard plane detection algorithm (called FUVAI) and evaluated the method on 50 freehand fetal US video scans. We compared FUVAI fetal biometric measurements with measurements made by five experienced sonographers at two time points separated by at least two weeks. Intra- and inter-observer variabilities were estimated. 
\newline
\textbf{Main Results:}
We found that automated fetal biometric measurements obtained by FUVAI were comparable to the measurements performed by experienced sonographers The observed differences in measurement values were within the range of inter- and intra-observer variability. Moreover, analysis has shown that these differences were not statistically significant when comparing any individual medical expert to our model. 
\newline
\textbf{Significance:} We argue that FUVAI has the potential to assist sonographers who perform fetal biometric measurements in clinical settings by providing them with suggestions regarding the best measuring frames, along with automated measurements. Moreover, FUVAI is able perform these tasks in just a few seconds, which is a huge difference compared to the average of six minutes taken by sonographers. This is significant, given the shortage of medical experts capable of interpreting fetal ultrasound images in numerous countries.
\end{abstract}

\vspace{2pc}
\noindent{\it Keywords}: Deep learning, Fetal imaging, Fetal biometric measurements, Fetal ultrasound video analysis, Inter- and intraobserver variability

\section{Introduction}

Fetal ultrasound (US) is an essential diagnostic tool used for assessing fetal growth and to detect abnormalities during pregnancy. Clinically, accurate fetal biometric measurements of head circumference (HC), biparietal diameter (BPD), abdomen circumference (AC) and femur length (FL), used to estimate gestational age (GA) and fetal weight (EFW), are crucial for proper delivery management \cite{Alberry}, \cite{Salomon}. Carrying out fetal body measurements is a task that requires following strict procedures which standardize the examination. 

The most important first step is the identification of \textit{standard planes} during the examination, which is a prerequisite for performing measurements based on standardized procedures. Standard planes are characterized by providing an optimal, standardized view of the examined structures based upon the presence of desired anatomical structures and their appropriate exposure \cite{March}.
Obtaining proper biometric measurements is subject to intra- and interoperator variabilities, and depends on both the correctness of standard plane acquisition and utilization of proper measuring technique \cite{SarrisVariability}.

Both tasks require substantial knowledge and experience on the part of the operator \cite{SharmaKnowledge}. Given the limited availability of expert sonographers, especially in underdeveloped countries \cite{shah}, \cite{van2019} , there is a need for an automated approach to standard plane identification in order to ensure correct measurement of fetal structures in video recordings of ultrasound examinations. Automated fetal US biometry may also help minimize variability for less experienced sonographers. Automation of fetal biometric measurements has been a field of interest for researchers and medical professionals since the early 1990’s \cite{Zador_1991}, \cite{Thomas_1991}. However, creating a computer program capable of mimicking the actions of an experienced sonographer requires solving two major issues, namely being able to assess whether the given frame satisfies the conditions of a standard plane, and creating accurate segmentations of fetal body parts that are measured during the ultrasound procedure. 

To automate fetal body part measurements researchers have applied computer-aided diagnosis methods, including advanced deep learning-based tools. Deep learning models gained popularity due to their high prediction accuracy, attaining human-level performance across different medical imaging applications, such as anatomical landmark detection in head CT \cite{ONeil}, pneumonia detection in chest X-ray \cite{CheXNet}, and head measurement in fetal ultrasound \cite{SinclairHead} to name just a few. Deep learning approaches have also been utilized in fetal standard plane classification \cite{Baumgartner}, \cite{BaumgarnerSP}, \cite{CaiSP}, \cite{BurgosSP}, \cite{LiangSP}, \cite{ChenSP}, of head \cite{SinclairHead}, \cite{Budd_2019Head}, \cite{Zeng_2021Head}, of abdomen \cite{RavishankarAbdomen}, \cite{Jang_2018Abdomen}, \cite{Kim_2018Abdomen}, \cite{Li_2020Abdomen} or both \cite{WuBoth} to improve computer-aided fetal biometric measurements. There exists commercial software, embedded in ultrasound devices, e.g. SonoBiometry (General Electric Healthcare) \footnote{GE Versana Club - SonoBiometry Online: https://www.versanaclub.net/emea/sono-biometry} or BiometryAssist (Samsung Healthcare) \footnote{Samsung Medison’s Fetal Ultrasound Smart Workflow - https://www.samsunghealthcare.com/en/products/UltrasoundSystem} that assists the sonographer by measuring fetal structures on still frames chosen by the operator. Such software helps reduce the number of keystrokes by providing suggestions on caliper placement (in the case of biparietal diameter and femur length measurements) or by drawing ellipses that are used to measure head and abdominal circumference. 
This kind of proprietary software is a semi-automatic solution, as it still requires the operator to select the correct frame, based upon their expertise.
To our knowledge, only \cite{LiuFemur}, \cite{BanoAll} directly tackle the problem of classification and segmentation of all three body parts using a single neural network.

\begin{figure*}
    \centering
    \includegraphics[width=\textwidth]{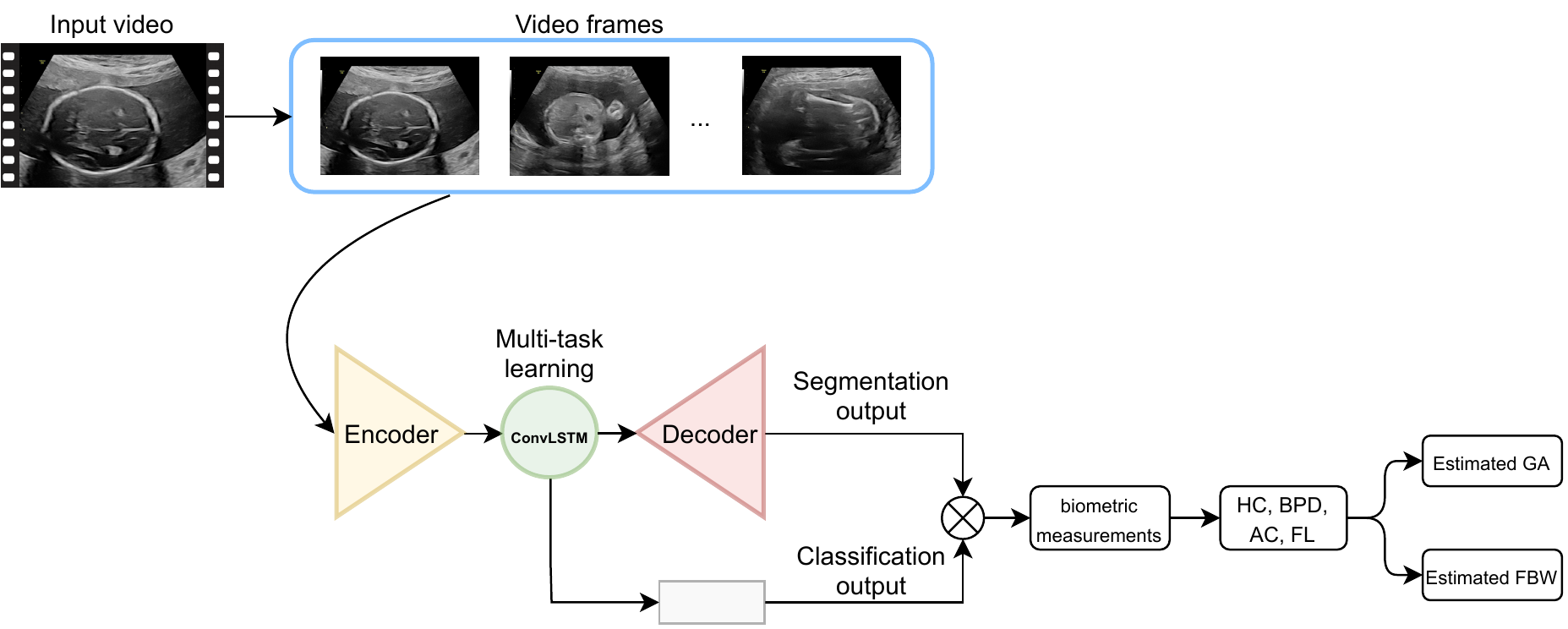}
    \caption{Overview of the proposed method. We use a fetal US video scan as input data. We train a multi-task neural network to learn $2D + t$ spatio-temporal features for simultaneous learning to classify, segment and measure fetal body parts. Next, based on automatic measurement of HC, BPD, AC and FL, we estimate gestational age and fetal weight.}
    \label{fig:fuvai}
\end{figure*}

Our method presented in Figure \ref{fig:fuvai} differs from \cite{LiuFemur} and \cite{BanoAll} in a few key aspects. Both of their models were trained on single-image frames which does not enable temporal analysis of fetal ultrasound recordings. Moreover, datasets used in both works contained only images of fetal body parts in their respective standard planes. This is an important limitation because ultrasound video recordings may contain numerous frames that are of no clinical use. This includes frames that either contain no fetal body parts or those that are not important in the examination, and also frames which depict the desired body parts but in a flawed view which renders them useless for performing biometric measurements. For this reason it is uncertain whether the results of their classifiers are useful for the purpose of choosing the best frames for clinical use, rather than simply distinguishing between individual categories of images.

To overcome these limitations, we propose a multi-task deep learning-based method for 2D+t spatio-temporal fetal US video scan analysis. Here, in addition to 2D images we rely on time which serves a {\em context}. This is similar to analysis of 3D images where neighboring slices are used to assist in classification tasks \cite{DingContext, YanContext}. The algorithm detects the best standard plane from the whole US video scan to automatically perform fetal biometric measurements on this plane. Estimations of gestational age and fetal weight are based on those measurements. Multi-task learning (MTL) aims to boost generalization and performance by simultaneously learning multiple related tasks. MTL can not only improve the performance of both tasks but also reduce overfitting through shared representations and speed up learning by leveraging auxiliary information \cite{ZhangMTL}. We also use a comprehensive fetal US video dataset acquired from 700 pregnant women between the 15th and 38th week of gestation to train the model. The used data allows the proposed method to better generalize upon the evaluation test set of fetal US video scans, with examinations from the beginning of the second trimester all the way to delivery. We compare measurements made by FUVAI with manual measurements by experienced sonographers. These experiments show that our proposed method has the potential to become an auxiliary tool for fetal biometric measurement in routine fetal ultrasound examinations in clinical settings. To the best of our knowledge, this is the first paper which compares the performance of human experts and deep learning-based methods in performing fetal biometric measurements on fetal head, abdomen and femur from US videos as opposed to single frames. The main novel contributions of this work are as follows:

\begin{itemize}
    \item We propose a multi-task deep learning-based method called FUVAI for $2D+t$ spatio-temporal fetal US video analysis. FUVAI is designed for automatic standard plane recognition and biometric measurement of fetal head, abdomen and femur directly in video recordings,
    \item We compare biometric measurements performed by the deep learning-based method with manual measurements by experienced sonographers using fetal US video recordings. Statistical analysis has proven FUVAI to be equally as good as experienced sonographers in both selection of the best standard planes and carrying out the actual measurements.
\end{itemize}
\noindent
The remaining sections are organized as follows. Section \ref{sec:methods} outlines our approach, datasets and a description of the proposed neural network. In Section \ref{sec:results} we describe the experiments and results, which are further discussed in Section \ref{sec:discussion}. Section \ref{sec:conclusions} concludes the paper.

\section{Methods}
\label{sec:methods}

\subsection{Fetal ultrasound datasets}
\begin{figure*}[t!]
\minipage{0.28\textwidth}
  \includegraphics[width=\linewidth]{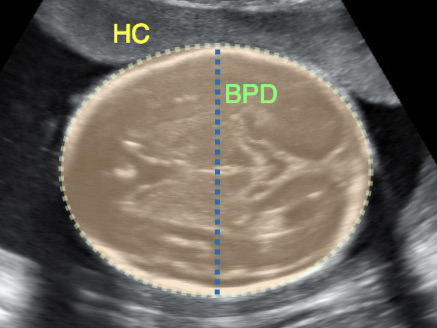}
\endminipage\hfill
\minipage{0.28\textwidth}
  \includegraphics[width=\linewidth]{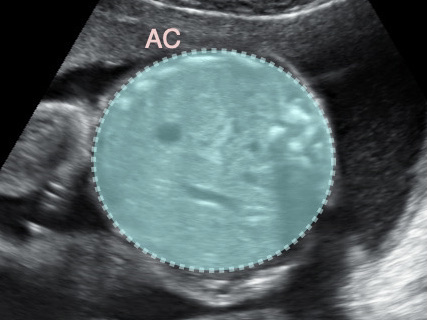}
\endminipage\hfill
\minipage{0.28\textwidth}%
  \includegraphics[width=\linewidth]{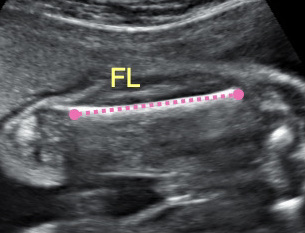}
\endminipage
\caption{Fetal biometry US standard planes: transventricular plane in the head (left), transabdominal plane in the abdomen (middle) and femur plane (right). Graphical definitions of HC, BPD, AC, and FL are shown.}
\label{fig:stplanes}
\end{figure*}

We use two datasets supplied by the University Centre of Mother and Child's Health of the Medical University of Warsaw to develop and evaluate our methods. Both fetal datasets were acquired following a pre-defined protocol pursuant to international standards approved by the International Society of Ultrasound in Obstetrics and Gynecology (ISUOG) \cite{Salomon}. The data comes from a single ultrasound device manufacturer (General Electric Healthcare) of several models with corresponding transabdominal transducers: S6 - RAB2-6-RS, S8 - RAB6-RS , P8 - RAB2-5-RS, E8 - RAB4-9-D, and E10 - RAB6-D. Prior to usage, both datasets were thoroughly anonymized in accordance with the ethical standards listed in the Declaration of Helsinki. Ethical approval was granted by the Ethics Committee of the Medical University of Warsaw. Both datasets consist of video recordings stored in DICOM file format captured in two resolutions: i) 975 $\times$ 742 pixels and ii) 1100 $\times$ 960 pixels. To ensure no sensitive information is present in the DICOM files, all metadata containing personal details is deleted. Next, the DICOM video files are converted to a series of images in the PNG (Portable Network Graphics) image file format, which is necessary for further training and evaluating the performance of our neural network. During this conversion, we remove sections of images that contain personal details and irrelevant information such as device settings, scale, etc., using a script that masks out the unwanted sections with black pixels. All frames were resized to $224 \times 224$ pixels without preserving the aspect ratio and normalized to 0-1 range. Examples of our annotations are presented in Figure \ref{fig:stplanes}. 

\noindent

\subsubsection{FUVAI development dataset}

The first dataset, which is used for the development of our neural network consists of video recordings from 700 pregnancies ranging between the 15th and 38th week of gestation, captured during routine fetal ultrasound examinations. To create this dataset sonographers were instructed to record three short US video clips per patient depicting the fetal head, abdomen and femur respectively. Operators were instructed to include the standard plane in the clips, but a specific location of the standard plane within the clip was not required. Each of the clips consist of between 250 to 460 frames and includes images that can be divided into three categories:
\begin{itemize}
    \item Images where the examined structures are visible in the standard plane,
    \item Images that contain the examined structures but are less correct than the above and do not meet the requirements necessary for classification as standard plane,
    \item Miscellaneous images that are of no clinical use due to technical issues (e.g. being out of focus or blurry), and/or contain either no fetal body parts, or such that are not relevant for the performed fetal ultrasound examination. 
\end{itemize}

\begin{figure*}[t!]
  \centering
\begin{tabular}{cccc}
\includegraphics[width=3.5cm]{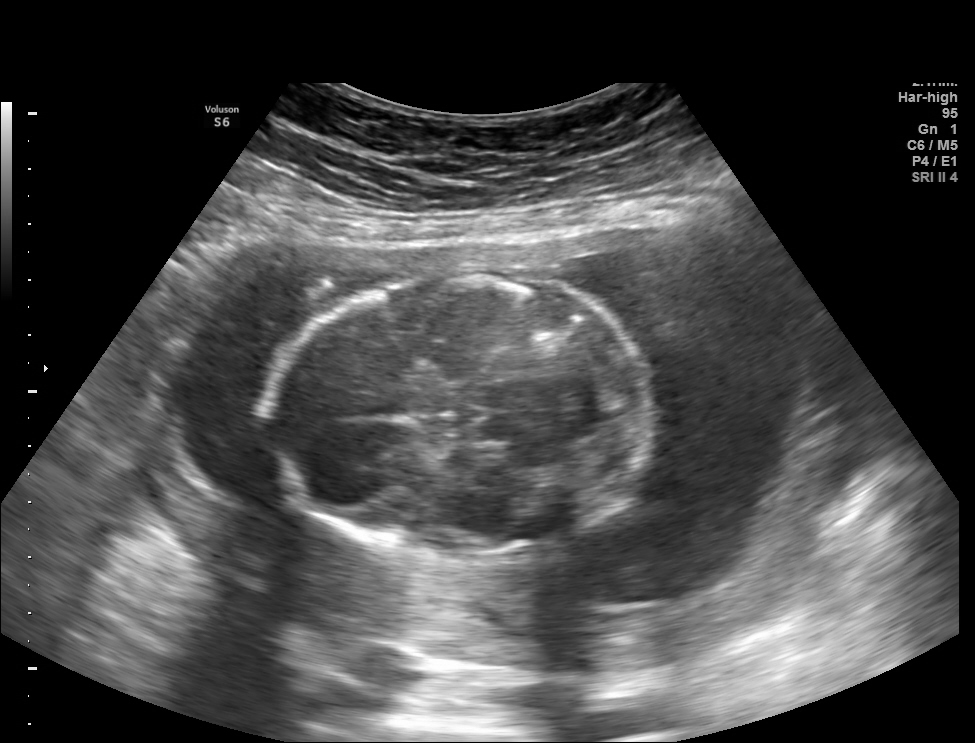}&
\includegraphics[width=3.5cm]{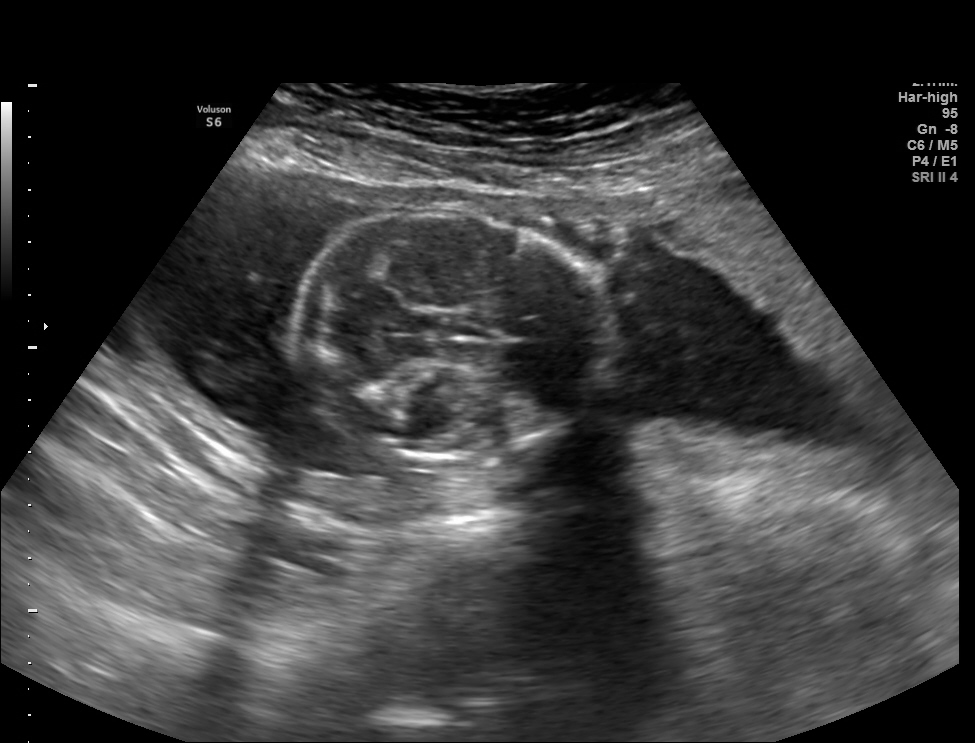}&
\includegraphics[width=3.5cm]{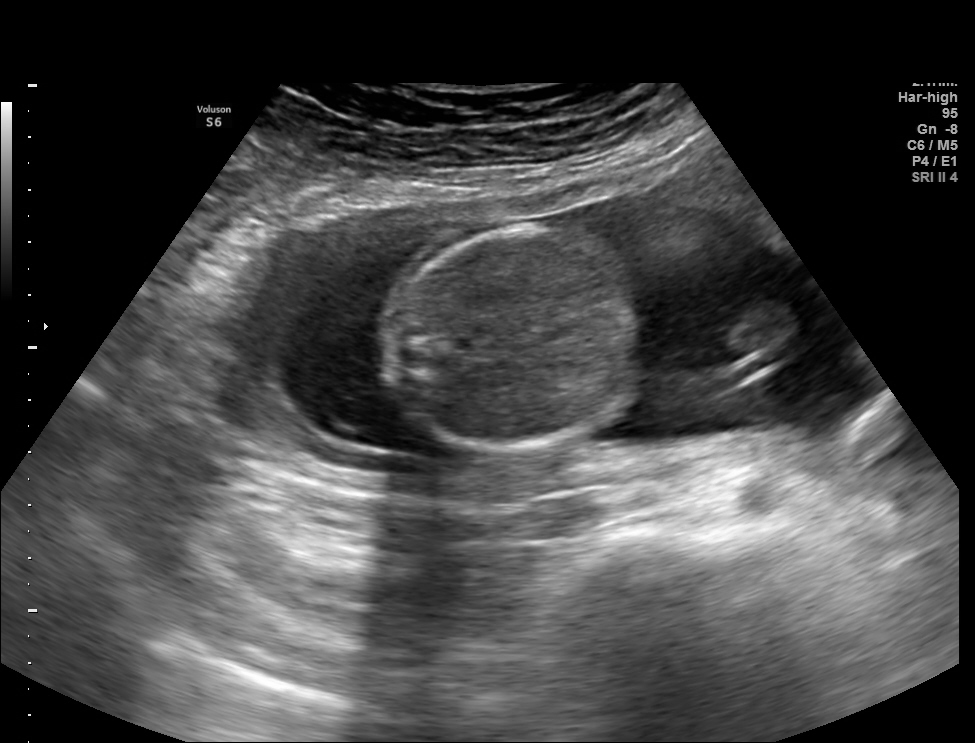}& 
\includegraphics[width=3.5cm]{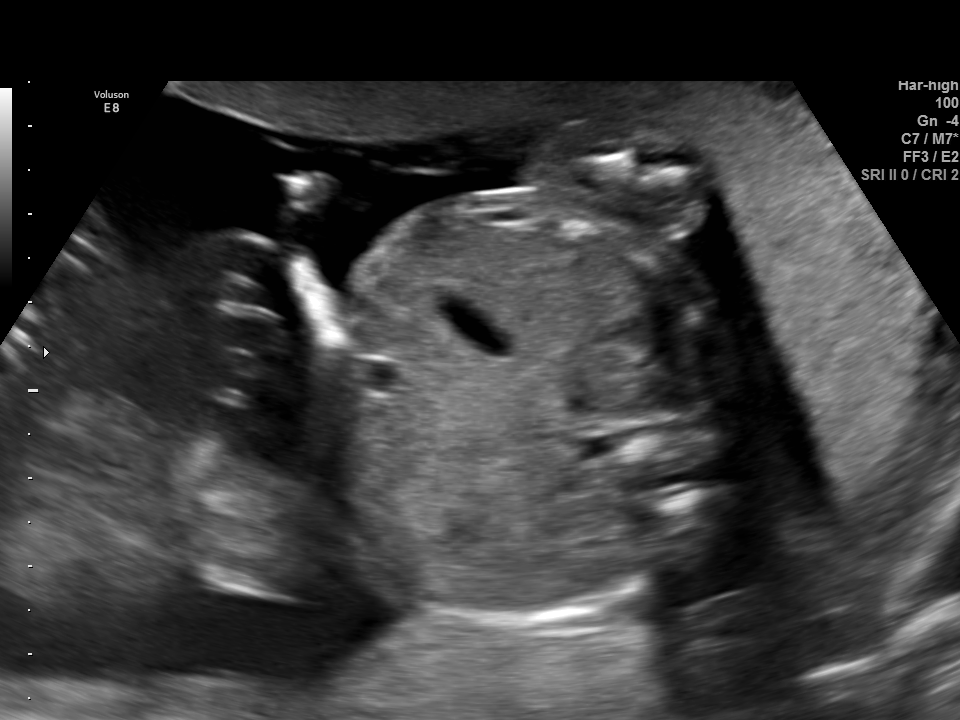} \\
\includegraphics[width=3.5cm]{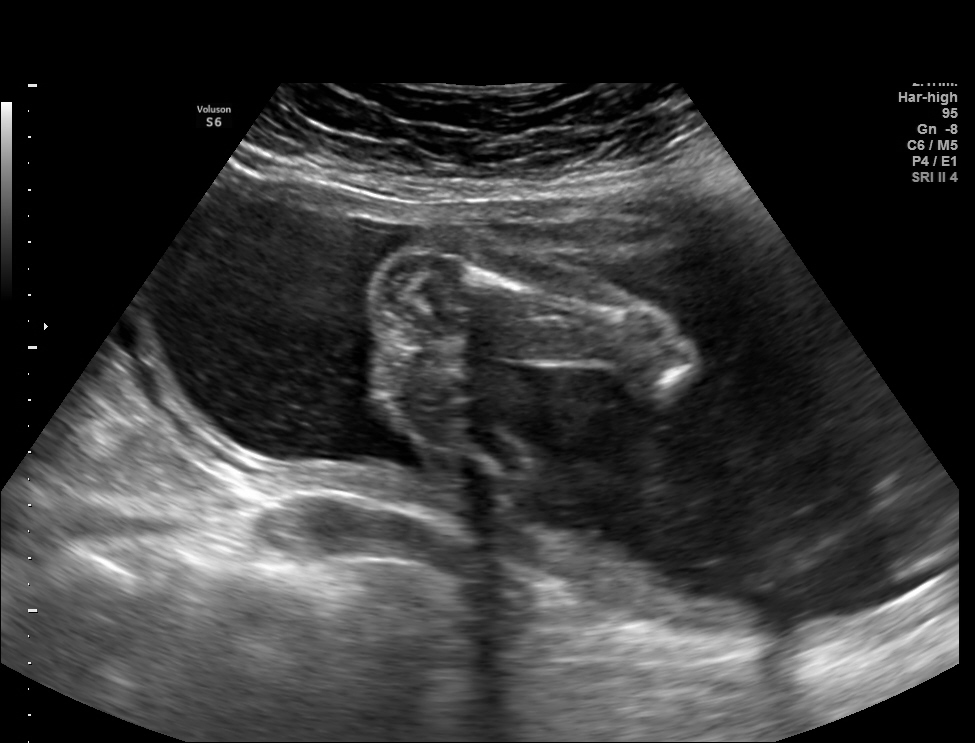}&
\includegraphics[width=3.5cm]{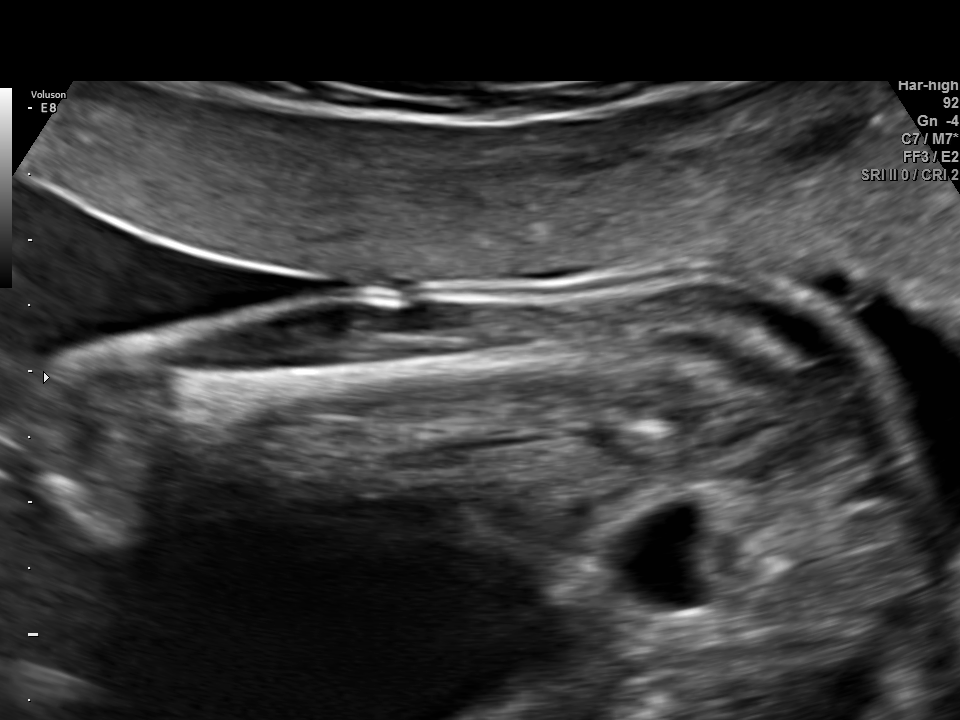}&
\includegraphics[width=3.5cm]{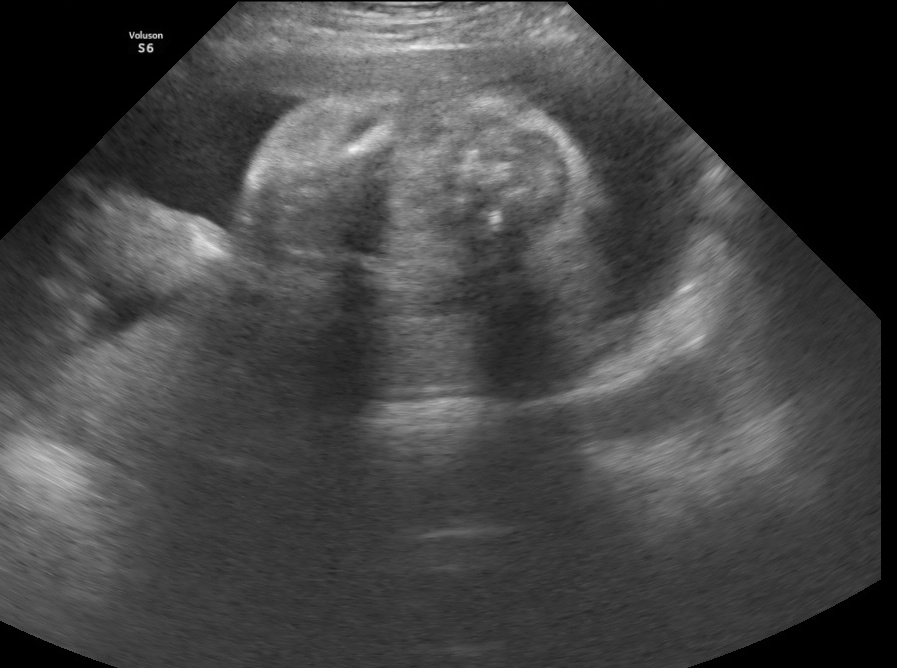}&
\includegraphics[width=3.5cm]{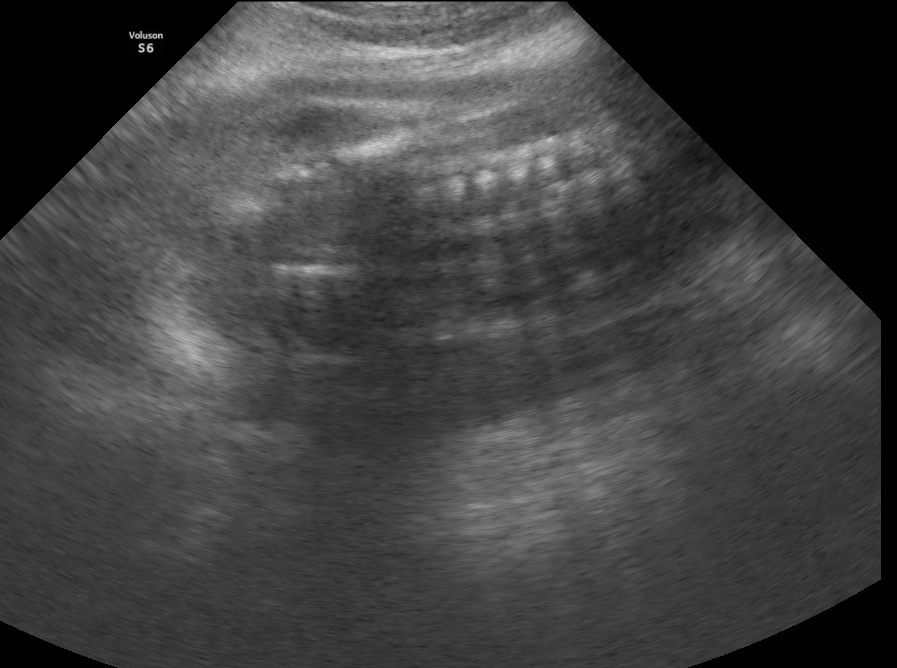} \\
\end{tabular}

    \caption{We show examples of the background class in fetal US video scans. From top left corner: two head sequences, two abdomen sequences, two femur sequences and two noise sequences.}
    \label{fig:background_planes}
\end{figure*}

All standard plane frames from the videos were labelled by experienced medical professionals in the form of graphical annotations, and provided along with numerical values of the measurements of head circumference, biparietal diameter, abdomen circumference and femur length taken during the examination. Additionally, frames that did not meet the criteria of standard planes, contained body parts other than head, abdomen or femur, or were technically flawed e.g. out of focus or blurry) were put together into a separate background category. Figure \ref{fig:background_planes} shows examples of the background class in the fetal US video scans.
The data was provided by six different expert sonographers with 40, 25, 20, 20, 15, and 8 years of experience. Importantly, they were not the same people as readers whose measurements were used in this study to compare the performance of medical experts versus our neural network, FUVAI.

We split the dataset by using 80\% of its contents for training and 20\% for testing. The training set is comprised of 32215 images labeled as heads, 26403 abdomens, 3706 femurs, and 211951 frames labeled as background, respectively. The test set consists of the remaining 7250 images labeled as heads, 6580 abdomens, 720 femurs and 42451 backgrounds.

\subsubsection{Freehand video test set}
In order to evaluate the performance of our FUVAI model against medical professionals, we designed a second dataset, consisting of 50 videos recorded during routine ultrasound examinations of women between the 19th and 38th week of pregnancy. The recordings are of different patients than those that were used to train our neural network. Sonographers who performed the examinations were instructed to record 1-2 minute videos (depending on their preference) during which fetal head, abdomen and femur standard planes were captured. Depending on the recording time and ultrasound machine, recordings consist of between 500 and 1900 frames. While recording, sonographers did not freeze the video to perform any measurements. The videos were recorded for the sole purpose of comparing the performance of our neural network against highly trained professionals.

\subsection{Automated measurements using multi-task neural network}
\label{sec:multitask_nn}

The network is designed to perform two tasks: segmentation and classification of fetal body parts which is subsequently used to perform measurements, and standard plane classification. We detail the design of this network in the next section.

\subsubsection{Network architecture and model training}

\begin{figure*}
    \centering
    \includegraphics[width=17cm]{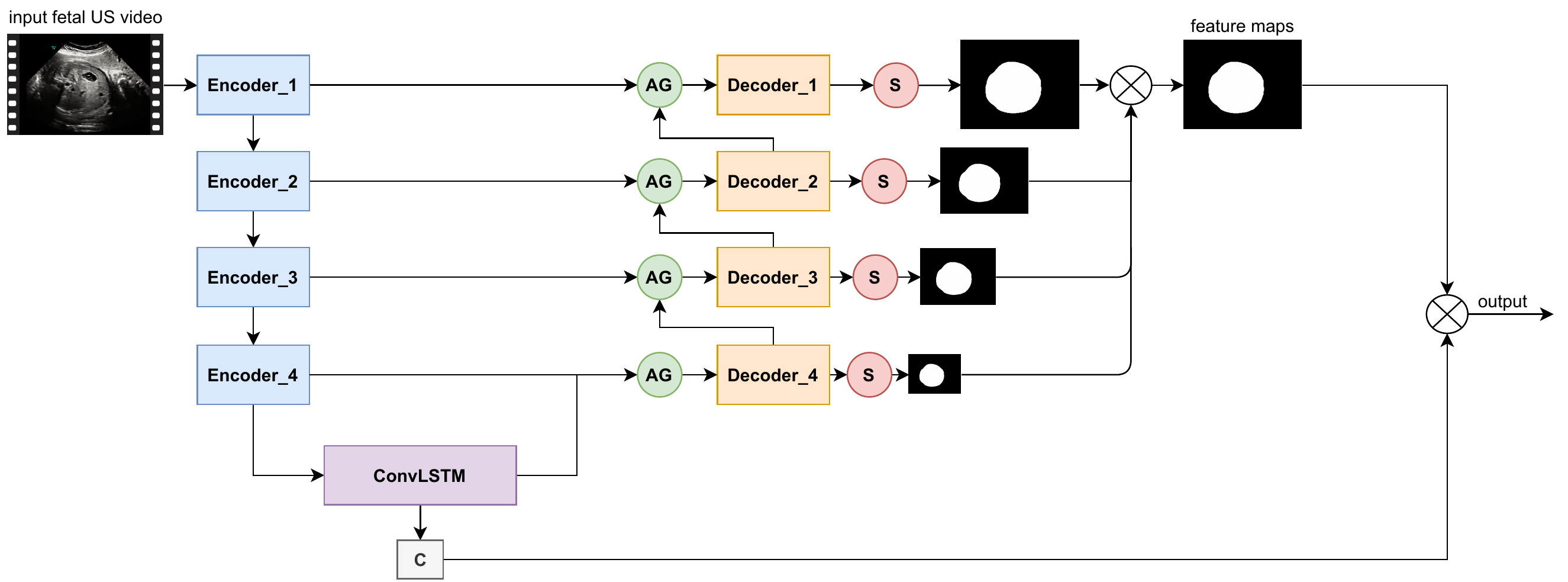}
    \caption{Overview of the proposed neural network. We use a multi-task neural network for $2D+t$ spatio-temporal feature analysis to simultaneously localize, classify and measure fetal body parts. The encoder part extracts spatial US image features and forwards them to the ConvLSTM cell to learn temporal features. We concatenate features from the encoder part via skip connections and attention gate (AG) in the decoder part. The skip connections carry only the spatial information (no temporal information) to the encoder part of the network. We scale output from each decoder by 1x1 convolutional layer (S) to the output size. Ultimately, we sum up features from each decoder block. For classification (C), we use spatio-temporal features to classify each fetal body part.}
    \label{fig:neuralnetwork}
\end{figure*}

Following \cite{Mehta}, \cite{wangsimultaneously} we use an encoder-decoder U-Net-based multi-task convolutional neural network architecture for joint segmentation and classification of fetal body parts in fetal ultrasound video scans. We extend the original U-Net implementation where each block consists of the following order: Conv3x3-BatchNorm-ReLU-Conv3x3-BatchNorm-ReLU-Dropout2D with $p=0.2$. After each block of the encoder part, we apply the Max Pooling layer with a kernel size of $2 \times 2$ and stride = 2. The number of feature maps in the input layer is equal to n = 64. The remainder of the eight convolutional blocks consists of 2n-4n-8n-16n-8n-4n-2n-n feature maps. We use encoder blocks to obtain high-level fetal US 2D spatial feature representations. The skip connections carry only the spatial information (no temporal information) to the encoder part of the network. The encoder's output (2D spatial information) is fed to the ConvLSTM-based \cite{ConvLSTM} bottleneck. The ConvLSTM cell is able to model 2D spatio-temporal image sequences by encoding their 2D spatial feature representation as temporal feature representation. Modeling $2D+t$ spatio-temporal image sequences by ConvLSTM cell effectively improves performance in both segmentation and classification \cite{ZhangConvLSTM}. We employ the attention gate mechanism \cite{AttentionUNet} to implicitly learn to suppress irrelevant regions in an input video sequence while highlighting the salient features of the target region of interest. The attention gate mechanism helps exploit local information to efficiently localize objects (i.e. fetal body parts) and improve prediction performance. Every encoder block forwards its output feature maps to the decoder part while concatenating them with an attention gate. To improve the performance of binary prediction feature maps, we employ deep supervision to connect the lower and higher scale levels of each decoder feature, creating what \cite{CascadedChen} call a stacked module. Multi-scale feature maps help encode both global and local contexts. We use a set of $3 \times 3$ 2D convolutional layers to up-sample the feature maps after each convolutional block. Thereafter, we combine the preceding high-level feature maps into an aggregate binary segmentation map. For the classification branch, we apply Adaptive Average Pooling 2D and Dropout2D with $p=0.4$ as ConvLSTM output before a Fully Connected layer with $14 \times 14 \times 16n$ feature maps on the output to assign video frames to one of the following classes: fetal head, abdomen, femur or background, at the frame level. Figure \ref{fig:neuralnetwork} shows the proposed multi-task learning method called FUVAI for $2D+t$ spatio-temporal fetal ultrasound scan video analysis. For more details of the model, please refer to our GitHub repository (https://github.com/SanoScience/FUVAI). 

We resize the input size of training images to 224 $\times$ 224 pixels image size and train our model until convergence over 100 epochs, with a batch size of 16, an initial learning rate of $10^{-4}$ and a weight decay factor of $10^{-4}$. To minimize the loss function, we set Adam as the optimiser. To prevent overfitting, we apply various data augmentation techniques. During training, we perform the following transformations: rotation between -15 and 15 degrees, contrast and brightness manipulation, as well as horizontal and vertical flipping. Each augmentation has a 50\% chance of being applied to each image during each epoch. We also apply a shuffled sampler. 

As the loss function, we use the sum of dice $L_{dice}$ and entropy $L_{CE}$ losses: 


\begin{center}
\begin{equation}
    L_{dice} \ =\ 1 - \frac{2 \sum_{i}^{N} p_{i}g_{i} + \epsilon}{\sum_{i}^{N}p_{i}^2 + \sum_{i}^{N}g_{i}^2 + \epsilon},    
\end{equation}
\end{center}
where $p_{i}$ is the prediction pixel value and $g_{i}$ is the ground truth pixel value. $\epsilon$ is a small number used to avoid calculating the $\log$ of 0.0, and ii) Cross-Entropy loss:
\begin{center}
\begin{equation}
    L_{CE} = - \sum_{i=1}^{n} t_{i}\log(p_{i}),
\end{equation}
\end{center}
where $t_{i}$ is the true label and $p_{i}$ is the softmax probability for the $i^{th}$ class.

We train our neural network on a workstation equipped with an AMD FX-8320@3.5Ghz CPU and NVIDIA Titan RTX 24GB GPU with CUDA 11.0. We use the PyTorch \cite{PyTorch} deep learning library for implementation of our model. The scripts and weights of the trained model are available on GitHub (https://github.com/SanoScience/FUVAI).

\subsection{Extraction of biometric measurements from network output}
The raw output of our neural network, FUVAI, consists of two components: segmentation and classification score. The following two sections describe how they are used in order to obtain meaningful biometric measurements from fetal US video recordings.

\subsubsection{Biometric measurements}

\begin{figure*}[t!]
    \centering
    \includegraphics[width=14cm]{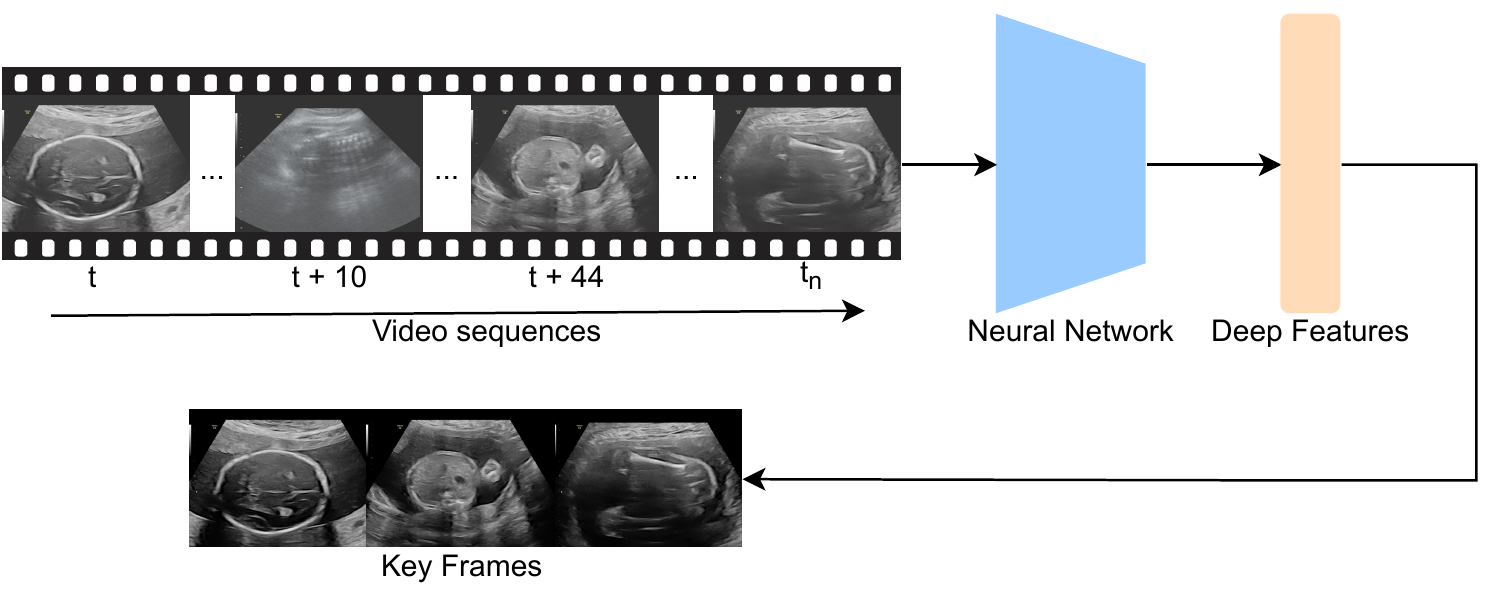}
    \caption{Overview of the standard plane detection algorithm. As input, we use a freehand fetal US video scan with various number of frames. Spatio-temporal features are computed through a neural network. Standard planes for each of the measured fetal body parts are obtained. In our work for each video up to three standard planes were detected.}
    \label{fig:my_label}
\end{figure*}

\label{sec:output_processing}
Since the segmentation output of our neural network takes the form of a $224 \times 224$ pixel binary mask, we first resize it to the match the size of the input image using bilinear interpolation. Next, we apply binary thresholding $p = 0.6$ and perform erosion followed by dilation, using a $5 \times 5$ cross-shaped structuring element. This ensures that the predicted masks are denoised. Finally, we use a median blur filter with a $13 \times 13$ kernel size to smooth the edges of the segmentations. Depending on the body part, we use different methods to obtain adequate measurements. For head and abdomen circumference measurement, we begin by finding the contours of the segmentation output. Next, we use Ramer-Douglas-Peucker approximation \cite{RamerBM}, \cite{DouglasPeuckerBM} and fit the ellipse to the postprocessed segmentation output using the direct least square method \cite{Fitzgibbon} We calculate the circumference of the fitted ellipse and store it. Additionally, to acquire the measurement of BPD, we store the length of the short axis of the ellipse fitted to the head \cite{BPD}. This implies that the values of BPD measurements taken by FUVAI are obtained similarly to the outer-outer measuring method. To obtain FL, we fit a rectangular bounding box to the contours of the segmentation. Next, we store the length of the fitted rectangle. Finally, we convert all measurements obtained in pixels to values in centimetres. We do this by multiplying the number of pixels by pixel size, an attribute that encodes the physical distance between centres of pixels, stored in DICOM  metadata.

\subsubsection{Selection of best frames from the video}

The raw network output consists of two components: segmentation output and classification score. Frames that have a high classification score (greater than 0.9) in one of the three categories -- head, abdomen of femur -- are marked as meeting the criteria of standard planes. Biometric measurements are performed for every standard plane frame as described in the above section \ref{sec:output_processing}. The obtained values are stored together with the frame indices. Once the entire recording is analyzed, best frames are selected. The best frame containing the femur is chosen by calculating a weighted average of the classification score and measurement value and selecting the frame for which this score is highest. Best head and abdomen frames are chosen similarly, but an additional computation is performed to compare the areas of the fitted ellipse and the raw segmentation output of the neural network. The calculated similarity score is used as the third component of the weighted average. Higher congruence of the ellipse and segmentation areas enables us to promote frames for which the segmentation is closest to the desired elliptical shape, and reject frames whose segmentations are irregular.

\subsubsection{Gestational age and fetal weight estimation}

Measurements of each of the body part for which adequate frames are present in the recording are cross-checked with WHO fetal growth charts, enabling assessment of fetal development against the background of population norms \cite{Kiserud}. For cases when head, abdomen and femur standard planes are all detected, $GA$ and $BW$ are estimated with the use of formulae \ref{eq:ga} and \ref{eq:bw}. If one or more body parts in standard plane views is not found, estimations of weight and gestational age are not performed. 

We calculate gestational age and fetal weight based on measurements of head circumference, abdomen circumference, biparietal diameter and femur length obtained by automatic measurements, using the following equation \cite{GA_estimation}: 
\begin{equation}
    \begin{split}
    GA &= 10.6 - 0.168 \times BPD + 0.045 \times HC \\ 
    &+ 0.03 \times AC + 0.058 \times FL \\
    &+ 0.002 \times BPD^2 + 0.002 \times FL^2 \\
    &+ 0.0005 \times (BPD \times AC) - 0.005 \times (BPD \times FL) \\ 
    &- 0.0002 \times (HC \times AC) + 0.0008 \times (HC \times FL) \\
    &+ 0.0005 \times (AC \times FL)
    \end{split}
    \label{eq:ga}
\end{equation}
Estimation of fetal weight relies on the Hadlock III formula \cite{Hadlock}  recommended by the World Health Organization:

\begin{equation}
    \begin{split}
        \log_{10} BW &= (1.326 - 0.00326 \times AC \times FL \\
        &+ 0.0107 \times HC + 0.0438 \times AC \\ 
        &+ 0.158 \times FL),
    \end{split}
    \label{eq:bw}    
\end{equation}
where HC, AC, FL, and BPD are head circumference, abdomen circumference, femur length and biparietal diameter respectively, as previously defined. They are expressed in centimeters. If standard planes corresponding to some of the measurements are not found and automated measurement is not available, $GA$ and/or $BW$ are not computed. 

\subsection{Evaluation metrics}
\subsubsection{Segmentation} To estimate segmentation performance, we use the following:

\begin{enumerate}
    \item Jaccard index, also known as Intersection over Union (IoU):
    \begin{equation}
    J(A,B) \ =\ \frac{|A\cap B|}{|A\cup B|},
    \label{eq:jaccard}
\end{equation}
    \item Dice coefficient (DSC), also known as Sorensen-Dice coefficient or F1 score:
    \begin{equation}
    DSC \ =\ 2 \times \frac{|A \cap B|}{|A| + |B|}.
    \label{eq:dice}
    \end{equation}
\end{enumerate}

\subsubsection{Classification} To measure the performance of the classification case, we employ the following metrics: 
\begin{enumerate}
    \item Accuracy:
    \begin{equation}
    Accuracy \ =\ \frac{TP + TN}{TP + FP + TN + FN},
    \label{eq:accuracy}
    \end{equation}
    \item Precision:
    \begin{equation}
    Precision \ =\ \frac{TP}{TP + FP},
    \label{eq:precision}
    \end{equation}
    \item Recall:
    \begin{equation}
    Recall \ =\ \frac{TP}{TP + FN},
    \label{eq:recall}
    \end{equation}
    \item F1 score:
    \begin{equation}
    F1 \ =\ 2 \times  \frac{Precision \times Recall}{Precision + Recall},
    \label{eq:f1}
    \end{equation}
\end{enumerate}
where TP, TN, FP, FN stand for true positive, true negative, false positive and false negative cases respectively. For multi-class classification, the values of TP, TN, FP and FN are computed for binary classification tasks of one class vs. others.

\subsection{State-of-the-art CNN architectures}

To compare the method developed in this work with other CNN networks, we implemented four state-of-the-art convolutional neural network architectures for both segmentation and classification of fetal body parts. For all deep learning methods, we used the Adam optimizer with an initial learning rate and weight decay of 0.0001. The learning rate was reduced by half every 50 epochs, once the validation loss stopped decreasing. In all experiments we used the sum of Dice loss and Cross-entropy loss as our loss function. To prevent overfitting, we applied early stops in the training phase if there was no improvement in validation loss after 30 epochs. Additionally, we used various training data augmentation methods (e.g. random rotation, contrast and brightness manipulation, as well as horizontal and vertical flipping) on the fly during training. The input for all state-of-the-art models were identical (image resolution, size, 2D image augmentations) as the input to FUVAI except the time domain as the state-of-the-art models can process only 2D images. All compared methods were implemented in Python 3.8 using the PyTorch deep learning library. We used open-source code available on GitHub, including the original U-Net implementation \footnote{https://github.com/milesial/Pytorch-UNet}, Fully Convolutional Network (FCN) \footnote{https://github.com/wkentaro/pytorch-fcn}, and DeepLabV3 \footnote{https://github.com/chenxi116/DeepLabv3.pytorch}. We adopted those implementations for our use case and provide details at https://github.com/SanoScience/FUVAI.

\subsection{Expert reader measurements}

We compared fetal biometric measurements and resulting estimates of gestational age and fetal weight between FUVAI and experienced sonographers with a test set of 50 freehand fetal US videos. Five readers participated in the study, including three senior gynaecologists with 40, 25 and 15 years of experience since completion of residency training, and two junior gynaecologists with less than 5 years of experience since completion of residency training.

Readers had no access to any clinical information regarding patients. Intra- and interobserver variabilities were estimated. Each reader performed two measurements on all 50 cases with at least two weeks in between. No information about the first read was provided to observers prior to their second read. Fetal body part measurements were summarized separately for HC, BPD, AC, FL, and estimated GA and FW, and for the first and second readings. Inter- and intraobserver agreement was calculated by using the mean and standard deviation value for all measurements, along with the intraclass correlation coefficient (ICC). ICC was calculated as an aggregate value (for FUVAI and five experienced sonographers), for the first and the second reading respectively.

\section{Results}
\label{sec:results}

\subsection{Comparison with other network architectures}

\begin{table}[ht!]
\small\addtolength{\tabcolsep}{-5pt}
\centering
\caption{Comparison of segmentation and classification of fetal body parts -- head, abdomen, femur and background -- with state-of-the-art neural networks and FUVAI}
\label{tab:segclsresults}
{%
\begin{tabular}{l|c|c|c|c|c|c}
\textbf{Method} & \multicolumn{1}{l|}{\textbf{IoU}} & \multicolumn{1}{l|}{\textbf{Dice}} & \textbf{Acc} & \textbf{Precision} & \textbf{Recall} & \textbf{F1}  \\ \hline
U-Net (base) & 0.862±4.02  & 0.921±3.98 & - & - & - & - \\
DeepLabv3 & 0.851±4.12 & 0.912±4.04 & 0.922 & 0.91 & 0.89 & 0.90  \\
FCN-8s & 0.865±3.88 & 0.924±3.79 & 0.933 & 0.93 & 0.91 & 0.92  \\
FCN-32s & 0.872±3.58 & 0.932±3.52 & 0.935 & 0.93 & 0.91 & 0.92  \\
\textbf{FUVAI (ours)} & \textbf{0.905±3.12} & \textbf{0.962±3.02} & \textbf{0.975} & \textbf{0.96} & \textbf{0.97} & \textbf{0.96}
\end{tabular}
}
\end{table}

We compare FUVAI with four state-of-the-art methods for multi-task learning. In Table \ref{tab:segclsresults}, we show results of the following neural networks: U-Net, FCN-8s, FCN-32s \cite{fcn_nn} and DeepLabv3 \cite{deeplab_nn}. We evaluate our model on 57001 test images of fetal head (7250 images), abdomen (6580 images), femur (720 images) and background (42451 images), and summarize those results in Table~\ref{tab:segclsresults}. With FUVAI we obtain average values of 0.905 and 0.962 for IoU and Dice respectively. Average precision, recall and F1 score are 0.96, 0.97 and 0.96 respectively. Based on Table~\ref{tab:segclsresults} we find that the proposed system outperforms the state-of-the-art neural networks. A one-way ANOVA was performed to compare state-of-the-art neural networks with the FUVAI method, revealing statistically significant differences in mean IoU and Dice (p = 0.29).
Table \ref{tab:measurementresults} compares results of fetal head, abdomen and femur error measurement (in mm) against state-of-the-art neural networks, presenting mean values and standard deviations. The mean errors are 2.9, 3.8 and 0.8 mm for fetal head circumference, abdomen circumference and femur length respectively. Comparison of clinical tests with predicted measurements shows that errors are lower than ±15\%, which is considered acceptable in clinical practice \cite{SarrisVariability}.

\begin{table}[]
\centering
\caption{Comparison of measurement error (in mm) of fetal body parts -- head, abdomen and femur -- for state-of-the-art neural networks and FUVAI.}
\label{tab:measurementresults}
{%
\begin{tabular}{l|c|c|c}
\textbf{Method} & \multicolumn{1}{l|}{\textbf{HC}} & \multicolumn{1}{l|}{\textbf{AC}} & \textbf{FL}  \\ \hline
U-Net (base) & 4.5±3.2  & 5.4±3.1 & 1.5±1.4 \\
DeepLabv3 & 4.8±3.4 & 5.5±3.4 & 1.5±1.3  \\
FCN-8s & 4.7±3.1 & 5.3±3.3 & 1.6±1.2  \\
FCN-32s & 3.9±2.8 & 4.9±3.2 & 1.2±0.8  \\
\textbf{FUVAI (ours)} & \textbf{2.9±1.2} & \textbf{3.8±3.0} & \textbf{0.8±1.2}
\end{tabular}
}
\end{table}

\subsection{Comparison between FUVAI and expert readers}

Table \ref{tab:meanmeasurements} shows descriptive statistics for the 50 freehand fetal ultrasound video scans. We computed mean measurement and standard deviation values of fetal body parts for FUVAI and for five experienced sonographers (ES1-ES5). For the second reading, we included the mean of absolute difference in measured values compared to the first reading. We note that FUVAI has similar performance and operates within the range of human-level error. Note that for FUVAI, the second reading is identical to the first reading due to the deterministic nature of neural network inference, resulting in intra-observer variability equal to zero, which is indicated in Table ~\ref{tab:meanmeasurements} by dashes.

\begin{table*}[ht!]
\caption{Mean measurements and standard deviation of each reader for all measurements. In the second reading, we show mean absolute differences between the first and the second reading.}
\label{tab:meanmeasurements}
\resizebox{\textwidth}{!}{%
\begin{tabular}{|l|c|ccc|cccc|}
\cline{2-9}
\multicolumn{1}{l|}{} & \multicolumn{4}{c|}{\textbf{1st reading}}                                                                  & \multicolumn{4}{c|}{\textbf{2nd reading}}                                                                                 \\ \hline
\textbf{Reader} & \textbf{HC [cm]} & \textbf{BPD [cm]}                & \textbf{AC [cm]}                  & \textbf{FL [cm]} & \textbf{HC [cm]}                 & \textbf{BPD [cm]}                & \textbf{AC [cm]}                 & \textbf{FL [cm]} \\ \hline
\textbf{FUVAI}  & 27.04 ± 5.27     & \multicolumn{1}{c|}{7.36 ± 1.47} & \multicolumn{1}{c|}{24.84 ± 5.38} & 5.25 ± 1.14      & \multicolumn{1}{c|}{-}           & \multicolumn{1}{c|}{-}           & \multicolumn{1}{c|}{-}           & -                \\
\textbf{ES1}    & 26.40 ± 5.42     & \multicolumn{1}{c|}{7.21 ± 1.53} & \multicolumn{1}{c|}{24.23 ± 5.40}  & 5.20 ± 1.13      & \multicolumn{1}{c|}{0.17 ± 0.13} & \multicolumn{1}{c|}{0.08 ± 0.04} & \multicolumn{1}{c|}{0.25 ± 0.17} & 0.04 ± 0.03      \\
\textbf{ES2}    & 26.58 ± 5.35     & \multicolumn{1}{c|}{7.22 ± 1.53} & \multicolumn{1}{c|}{24.40 ± 5.40}  & 5.21 ± 1.14      & \multicolumn{1}{c|}{0.23 ± 0.18} & \multicolumn{1}{c|}{0.08 ± 0.04} & \multicolumn{1}{c|}{0.23 ± 0.14} & 0.07 ± 0.08      \\
\textbf{ES3}    & 26.94 ± 5.27     & \multicolumn{1}{c|}{7.33 ± 1.55} & \multicolumn{1}{c|}{24.61 ± 5.36} & 5.26 ± 1.13      & \multicolumn{1}{c|}{0.25 ± 0.18} & \multicolumn{1}{c|}{0.08 ± 0.07} & \multicolumn{1}{c|}{0.29 ± 0.23} & 0.04 ± 0.04      \\
\textbf{ES4}    & 26.69 ± 5.32     & \multicolumn{1}{c|}{7.23 ± 1.52} & \multicolumn{1}{c|}{24.49 ± 5.34} & 5.21 ± 1.13      & \multicolumn{1}{c|}{0.29 ± 0.23} & \multicolumn{1}{c|}{0.07 ± 0.05} & \multicolumn{1}{c|}{0.27 ± 0.18} & 0.04 ± 0.04      \\
\textbf{ES5}    & 27.09 ± 5.36     & \multicolumn{1}{c|}{7.30 ± 1.54}  & \multicolumn{1}{c|}{24.85 ± 5.35} & 5.30 ± 1.11      & \multicolumn{1}{c|}{0.20 ± 0.16} & \multicolumn{1}{c|}{0.07 ± 0.05} & \multicolumn{1}{c|}{0.24 ± 0.17} & 0.03 ± 0.03     \\
\hline
\end{tabular}%
}
\end{table*}

In Table ~\ref{tab:fuvaiobservers}, we show the mean absolute error (MAE) between FUVAI and each of the five experienced sonographers (ES1-ES5) for both readings. We obtained MAE values of 1.04, 0.27, 1.06 and 0.20 cm for HC, BPD, AC and FL respectively. We estimated GA and fetal weight with MAE of 0.05 ± 0.01 week and 25 ± 5 g respectively. 

\begin{table}[ht!]
\caption{Mean absolute errors between FUVAI and observers in both readings.}
\centering
\label{tab:fuvaiobservers}
\begin{tabular}{|l|c|c|c|c|} 
\cline{2-5}
\multicolumn{1}{l|}{}                 & \multicolumn{4}{c|}{\textbf{FUVAI}}                                         \\ 
\hline
\multicolumn{1}{|c|}{\textbf{Reader}} & \textbf{HC [cm]} & \textbf{BPD [cm]} & \textbf{AC [cm]} & \textbf{FL [cm]}  \\ 
\hline
\textbf{ES1}                          & 1.07             & 0.25              & 1.13             & 0.16              \\
\textbf{ES2}                          & 1.02             & 0.25              & 1.06             & 0.20              \\
\textbf{ES3}                          & 1.03             & 0.30              & 0.99             & 0.19              \\
\textbf{ES4}                          & 0.99             & 0.27              & 1.06             & 0.21              \\
\textbf{ES5}                          & 1.10             & 0.29              & 1.10             & 0.22              \\
\textbf{Mean}                         & \textbf{1.04}    & \textbf{0.27}    & \textbf{1.06}   & \textbf{0.20}    \\
\hline
\end{tabular}
\end{table}

The overall inter- and intraobserver agreement is similar: for HC, BPD, AC and FL measurements, interobserver agreement rates are 0.974, 0.978, 0.963, 0.983 while intraobserver agreement rates are 0.972, 0.979, 0.961 and 0.978. A one-way ANOVA was performed to compare the measurements performed by the five readers (ES1-ES5) with the FUVAI method. ANOVA results are: F(5, 294) = 0.133, p = 0.985 for HC, F(5, 294) = 0.091, p = 0.993 for BPD, F(5, 294) = 0.11, p = 0.991 for AC and F(5, 294) = 0.052, p = 0.998 for FL respectively. A one-way ANOVA revealed that there was no statistically significant difference in mean measurement values between annotators (both human and automatic).

We performed further subanalysis based on reader experience and the specific trimester of pregnancy. Both inter- and intraobserver agreements differ depending on the reader experience level and trimester. We obtained ICCs between FUVAI and juniors of 0.982, 0.989, 0.985, 0.981 for HC, BPD, AC and FL respectively. and ICCs between FUVAI and seniors of 0.987, 0.991, 0.987, 0.986 for HC, BPD, AC and FL respectively. This shows that FUVAI results correlate better with seniors. For the 2nd and 3rd trimester of pregnancy the corresponding values are 0.982, 0.994, 0.980, 0.981 and 0.982, 0.995, 0.982, 0.983 for HC, BPD, AC and FL respectively. No statistically significant differences were detected between the second and third trimester of pregnancy.

\section{Discussion}
\label{sec:discussion}

We propose a novel multi-task encoder-decoder deep learning-based framework for fetal ultrasound video scan analysis and interpretation referred to as FUVAI. The success of the proposed method rests upon two factors. Compared with previous computer-aided methods, FUVAI is able to automatically analyze $2D+t$ spatio-temporal fetal ultrasound video scans simultaneously localizing standard planes, classifying and measuring fetal body parts. 

The ConvLSTM cell-based neural networks used here are efficient in encoding $2D+t$ spatio-temporal information and representation of features. Equally importantly, FUVAI takes advantage of the attention mechanism followed by multi-scale features in each decoder block, and achieves better accuracy in both segmentation and classification. The multi-scale feature information decoder is vital in the fetal body part segmentation task, given their considerable variations in terms of size, shape and location. The approach of relating segmentation performance to the amount of multi-scale feature representations is also true for 2D convolutional neural network-based methods, which rely solely on spatial features \cite{SinhaMultiScale}. For more details please refer to a separate ablation study which we published in a conference abstract~\cite{FetalNet}.

We have shown that human measurements differed, on average, by 1.04, 0.27, 1.06 and 0.20 cm for HC, BPD, AC and FL, compared with FUVAI (Table \ref{tab:fuvaiobservers}). These differences are consistent with interobserver variability documented by Sarris {\em et al.}~2012 \cite{SarrisVariability} where interobserver variability was reported as 0.99, 1.35, and 1.43 cm for HC, AC, and FL respectively. Moreover, Sinclair {\em et al.}~2018 \cite{SinclairHead} found interobserver MAE of 2.16 cm for HC and 0.59 cm for BPD and model-observer MAE 1.99 and 0.61 for HC and BPD respectively. 

\begin{figure}[ht!]
    \centering
\begin{tabular}{cc}
\includegraphics[width=6cm]{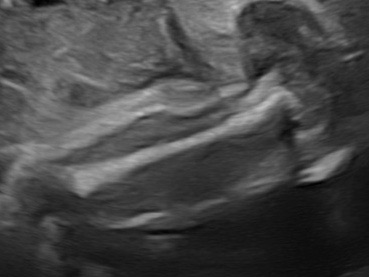}&
\includegraphics[width=6cm]{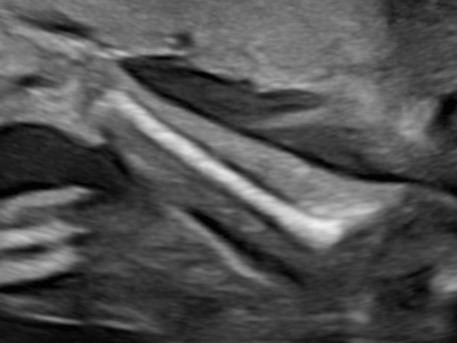}\\
\end{tabular}
    \caption{Femur standard planes missed by two of the readers, but correctly identified by FUVAI.}
    \label{fig:omitted}
\end{figure}

Obviously, FUVAI inference is deterministic and will always provide the same answer given the same input. It is important to note that considerable time is required for human experts to perform measurements -- on average, six minutes per movie, whereas FUVAI inference time is on the order of one second. Interestingly, FUVAI found two standard planes which were missed by two expert readers, and are shown in Figure~\ref{fig:omitted}. After the experiment, upon presentation of standard planes found by FUVAI (Figure~\ref{fig:omitted}), expert readers confirmed that they were indeed correct femur standard planes. 

The prevalence of the femur is naturally low. However, because of the high contrast of the femur we found that the model had no major difficulties in finding femur structures and decided against using any method to mitigate data imbalance, having in mind that such methods require additional hyperparameters which makes training more difficult and less robust.

In this work we consider a situation when the US examination is performed, stored in the Picture Archiving and Communication System (PACS) and read from there. However, this automatic approach can also be used in real time during routine fetal ultrasound examinations in which case it can help the operator identify standard planes and even automatically perform biometry while performing the scan. This, however, would require the software to be installed on the ultrasound device and integrated in the workflow. With regard to future work, the proposed method seems an attractive add-on that can be used on point-of-care portable devices, for example in underdeveloped countries. This, however, requires further validation with data coming from such devices -- which is likely to be of poorer quality than the data used in this work.

\begin{figure}[ht!]
    \centering
\begin{tabular}{cc}
\includegraphics[width=6cm]{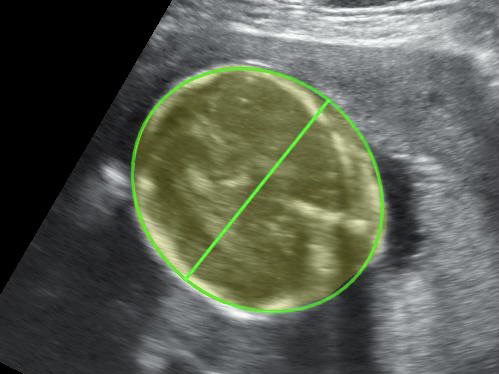}&
\includegraphics[width=6cm]{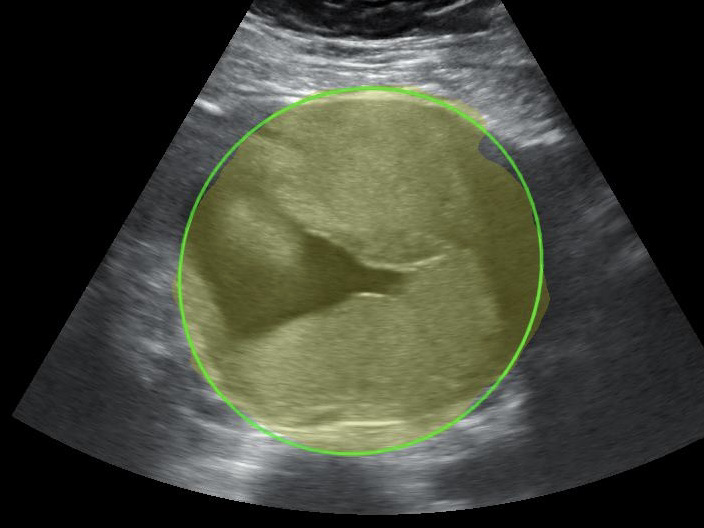}\\
\end{tabular}
    \caption{Two examples of errors made by FUVAI. Oversized prediction of fetal head (+ 2.5 cm) and misclassified standard plane of fetal abdomen respectively.}
    \label{fig:missclassified_error}
\end{figure}

Our study contains limitations. The method was trained on data from several models of a single ultrasound device manufacturer (General Electric Healthcare) and was acquired at a single institution. Therefore, it is uncertain if the findings are transferable to other manufacturers and institutions. Although we used a large training dataset, to our knowledge the largest in published literature, accuracy can likely be improved by using more training cases and more diverse training examples. The automatic method is not error-free (see examples in Figure~\ref{fig:missclassified_error}) and may be improved upon if a more diverse training set with more examples of images with shadowing/movement is used. Each image in the training set was annotated by a single annotator and it is possible that, if more annotators are used, the quality of inference will improve. We plan to add more annotators in the future. As for many other tasks in radiology considered for AI, the ground truth was defined subjectively by expert readers and no objective ground truth was available. A notable difference in the methodology of biparietal diameter measurement is present in this work. While sonographers who gathered data for both datasets use the inner-outer method of BPD measurement, FUVAI performs this measurement by calculating the length of the ellipses' short axes, which is more akin to the alternative outer-outer method. Both of these methods are clinically valid. Moreover, literature suggests that the differences between measurements obtained with the use of these methods are negligible \cite{BPD}. We aim to resolve this issue by developing a more specific method of BPD measurement.

\section{Conclusions}
\label{sec:conclusions}

We propose a multi-task deep learning-based framework for fetal ultrasound video scan analysis and interpretation referred to as FUVAI. The method is designed to process fetal US video scans to simultaneously localize standard planes in video sequences, classify and measure the fetal biometric parameters, and estimate gestational age and fetal weight. We demonstrated that the method achieves human-level performance, comparable to inter-rater agreement involving experienced sonographers. The method has the potential for use as a fetal biometry assistance tool that may be especially useful for less experienced personnel, and may save time when reading fetal ultrasounds. We provide details of the model and weights at https://github.com/SanoScience/FUVAI.

\section*{Acknowledgements}

The authors would like to thank the following medical sonographers for data, annotations and clinical expertise: Jan Klasa, MD; Bogusław Marinković, MD; Wojciech Górczewski, MD; Norbert Majewski, MD; Anita Smal-Obarska, MD and Robert Brawura-Biskupski-Samaha, MD, PhD. This publication is partly supported by the European Union’s Horizon 2020 research and innovation programme under grant agreement Sano No. 857533 and the International Research Agendas programme of the Foundation for Polish Science, co-financed by the European Union under the European Regional Development Fund. We would like to thank Piotr Nowakowski for his assistance with proofreading the manuscript.

\end{document}